\documentclass[%
 aip,
 amsmath,amssymb,
preprint,%
]{revtex4-1}

\usepackage{graphicx}
\usepackage{dcolumn}
\usepackage{bm}
\usepackage[utf8]{inputenc}
\usepackage[T1]{fontenc}
\usepackage{mathptmx}
\usepackage{xcolor}
\usepackage{url}
\usepackage{hyperref}
\usepackage[normalem]{ulem}
\usepackage{subfigure}

\hypersetup{
 colorlinks = true,
 allbordercolors = {white},
 allcolors = {blue},
 }
\usepackage[symbol]{footmisc}

\begin{document}

\title{Evolution of Perturbation in Quiescent Medium}

\author{Tapan K. Sengupta\footnote{Corresponding author.}}
\email{tksengupta@iitism.ac.in}
\affiliation{Department of Mechanical Engineering, IIT (ISM) Dhanbad, Jharkhand-826 004, India}

\author{Shivam K. Jha}
\affiliation{Department of Mechanical Engineering, IIT (ISM) Dhanbad, Jharkhand-826 004, India}

\author{Aditi Sengupta}
\affiliation{Department of Mechanical Engineering, IIT (ISM) Dhanbad, Jharkhand-826 004, India}

\author{Bhavna Joshi}
\affiliation{Department of Mechanical Engineering, IIT (ISM) Dhanbad, Jharkhand-826 004, India}

\author{Prasannabalaji Sundaram}
\affiliation{Department of Mechanical Engineering, IIT (ISM) Dhanbad, Jharkhand-826 004, India}

\date{\today}

\begin{abstract}
Here, the perturbation equation for a  dissipative medium is derived from the first principle from the linearized compressible Navier-Stokes equation without Stokes' hypothesis.  The dispersion relations of this generic governing equation are obtained for one and three-dimensional perturbation , which exhibit both the dispersive and dissipative nature of the perturbations traveling in a  dissipative medium, strictly depending upon the length scale. We specifically provide a theoretical cut-off wave number above which the perturbation equation represents diffusive and dissipative nature. Such behavior has not been reported before, as per the knowledge of the authors.  

\end{abstract}

\maketitle

\textbf{keywords:} perturbation equation; perturbation pressure;  dissipative medium; cut-off wavenumber\\
\\
Acoustics, as a branch of science, intrinsically deals with the propagation of signal (information) observed at one point to another closely related signal at another space-time location. Despite a long history of research on wave propagation in this context, there is no clear definition of waves \cite{Whitham74}.  The canonical wave equation was first described by D'Alembert \cite{DAlembert1750} as, 
\begin{equation} \label{Eq:utt}
    u_{tt} = c^2 u_{xx}
\end{equation}
 in the context of the one-dimensional transverse vibration of string in tension. The solution of Eq.~\eqref{Eq:utt}, subject to initial conditions, can be found in textbooks (see, e.g., Refs.~\onlinecite{Sengupta13, SenguptaBhum20}). This non-dissipative and non-dispersive (i.e. frequency-wavenumber independence) solution also sets a standard benchmark for developing and calibrating numerical methods in different branches of engineering and applied physics. 

Maxwell \cite{maxwell54, maxwell1865} obtained the wave equations for the electric field E, and the magnetic field B, with c as the speed of light (phase speed) in a medium of permeability $\mu_p$ and permittivity $\epsilon_p$ given by $c = 1/\sqrt{\mu_p \epsilon_p}$. An electromagnetic wave is transverse in nature, with E and B being perpendicular to wave propagation's direction. 

Some of the other physical phenomena governed by the partial differential equation (PDE) \eqref{Eq:utt} are listed in Mulloth et al. \cite{Mullothetal2015}. Among these are the use of classical wave equation in acoustics, Feynman \cite{Feynman65, Feynman69}; elastic wave propagation in solid mechanics \cite{Whitham74} relating applied strain and stress, with the longitudinal displacement $u$, given by Eq.~\eqref{Eq:utt} with $c^2 = E_0/\rho$, where $E_0$ is Young's modulus, and $\rho$ is the density of the medium. 

The present interest in information propagation as sound and flow perturbation arises from a desire to develop the common acoustic and fluid mechanics description from the first principle, for the unified description of disturbance propagation in a  dissipative medium for problems in the continuum. In the process of this development, a novel result is developed for scale-wise propagation of disturbances, following different physical mechanisms without requiring any restriction for the ensuing physical processes of convection and diffusion. 

For  problems in sound propagation, the compressible Navier-Stokes equation is mandatory to be used, with the disturbance treated as a small perturbation following the conservation of mass and momentum. In a quiescent, homogeneous medium, one can consider the equilibrium state with the following ansatz: disturbance velocity $\vec V '$ and disturbance density $\rho '$ develops with no mean motion ($\vec{ \bar V} = 0$) and a steady state for the unperturbed density prevails, i.e., $\left( \frac{\partial \bar \rho}{\partial t} = 0\right)$. 

Conservation of mass: For the overall perturbation field, this is given by,
\begin{equation} \label{Eq:CE}
    \frac{\partial \rho}{\partial t} + \nabla \cdot \rho \vec{V} = 0 
\end{equation}
The conservation of momentum equation without any body force is given by,  
\begin{eqnarray}
&&\rho \left( \frac{\partial \vec V}{\partial t} + \left( \vec V \cdot \nabla\right) \vec V \right)  = -\nabla p  \nonumber \\
&&+ \nabla \cdot \left( \lambda \left(\nabla \cdot \vec V \right) \bf{I} \right) + \nabla \cdot \left[ \mu \left( \nabla \vec V + \nabla \vec V ^T \right)\right]  \label{Eq:ME}
\end{eqnarray}

Here, $\bf{I}$ is an identity matrix with rank three. If one considers the acoustic signal as a small perturbation over the mean flow, then the velocity, the density, and the pressure can be expressed as a superposition of the unperturbed equilibrium state with the small perturbation given by, $\vec V = \vec{\bar V} + \epsilon \vec V ' ; ~\rho     = \bar \rho + \epsilon \rho ' ;~ p = \bar p + \epsilon p '$. 
%
Without any loss of generality, one can consider the propagation of the disturbances in a quiescent flow (i.e., $\vec {\bar V} = \vec 0$) in a homogeneous medium (i.e. constant $\bar \rho$), so that, 
\begin{equation} \label{Eq:04}
\vec V = \epsilon \vec V'.
\end{equation}
The $O(\epsilon)$ equation resulting from the conservation of mass equation, Eq.~\eqref{Eq:CE} yields, 
\begin{equation} \label{Eq:05}
    \frac{\partial \rho '}{\partial t} + \bar \rho  \nabla \cdot \vec{V}' = 0 
\end{equation}	Similarly, the $O(\epsilon)$ equation resulting from the conservation of momentum equation, Eq.~\eqref{Eq:ME} yields, 
\begin{equation} \label{Eq:06}
\bar \rho  \frac{\partial \vec V '}{\partial t}  = -\nabla p' + \nabla \cdot  \left( \lambda \left(\nabla \cdot \vec V'  \right)\bf{I} \right) + \nabla \cdot \left[ \mu \left( \nabla \vec V' + \nabla \vec V^{' T} \right)\right]
\end{equation}
From Eq.~\eqref{Eq:05}: $\nabla \cdot \vec V ' = -\frac{1}{\bar \rho} \frac{\partial \rho '}{\partial t}$ and differentiating this with respect to time yields, 
\begin{equation} \label{Eq:07}
\frac{\partial}{\partial t} \left(  \nabla \cdot \vec V '\right) = -\frac{1}{\bar \rho} \frac{\partial^2 \rho'}{\partial t ^2}
\end{equation}
Taking divergence of Eq.~\eqref{Eq:06}, one gets
\begin{equation} \label{Eq:08}
\bar \rho \frac{\partial}{\partial t} \left(  \nabla \cdot \vec V '\right)  = -\nabla^2 p' + \lambda \nabla^2 \left( \nabla \cdot \vec V' \right) +  2\mu  \nabla^2 \left(\nabla \cdot \vec V'\right)
\end{equation}
From Eq.~\eqref{Eq:07} one gets, 
\begin{equation} \label{Eq:09}
-\frac{\partial^2 \rho'}{\partial t^2}  = -\nabla^2 p' - \left( \lambda + 2 \mu\right) \nabla^2 \left( \frac{1}{\bar \rho} \frac{\partial \rho'}{\partial t} \right) 
\end{equation}
From the polytropic relation one gets, 
\begin{equation} \label{Eq:10}
	\frac{\partial \rho'}{\partial t} = \frac{1}{c^2}\frac{\partial p'}{\partial t}
\end{equation}
Therefore, eliminating $\rho'$ using this relation in Eq.~\eqref{Eq:09}, one gets
\begin{equation} \label{Eq:11}
\frac{1}{c^2}\frac{\partial^2 p'}{\partial t^2}  =\nabla^2 p' + \frac{\lambda + 2 \mu}{\bar \rho c^2}  \nabla^2 \left( \frac{\partial p'}{\partial t} \right) 
\end{equation}
\noindent which can be further simplified as,
\begin{equation} \label{Eq:12}
\frac{\partial^2 p'}{\partial t^2}  = c^2\nabla^2 p' + \nu_l \frac{\partial }{\partial t} \nabla^2  p'
\end{equation}
\noindent where the generalized viscosity is defined as, $\nu_l = \frac{\lambda + 2 \mu}{\bar \rho}$. Hence, Eq.~\eqref{Eq:12} implies that the Stokes' hypothesis \cite{stokes1851} is not used and one can incorporate the action of first and second coefficient of viscosities, as contributed by the bulk viscosity as an augmented loss term. In other words, effects of bulk action during the propagation of the signal as compression and dilation waves is kept under consideration. This can be also used for the generic acoustic equation in a  dissipative medium.

\begin{figure*}
\centering
\includegraphics[width=0.9\textwidth]{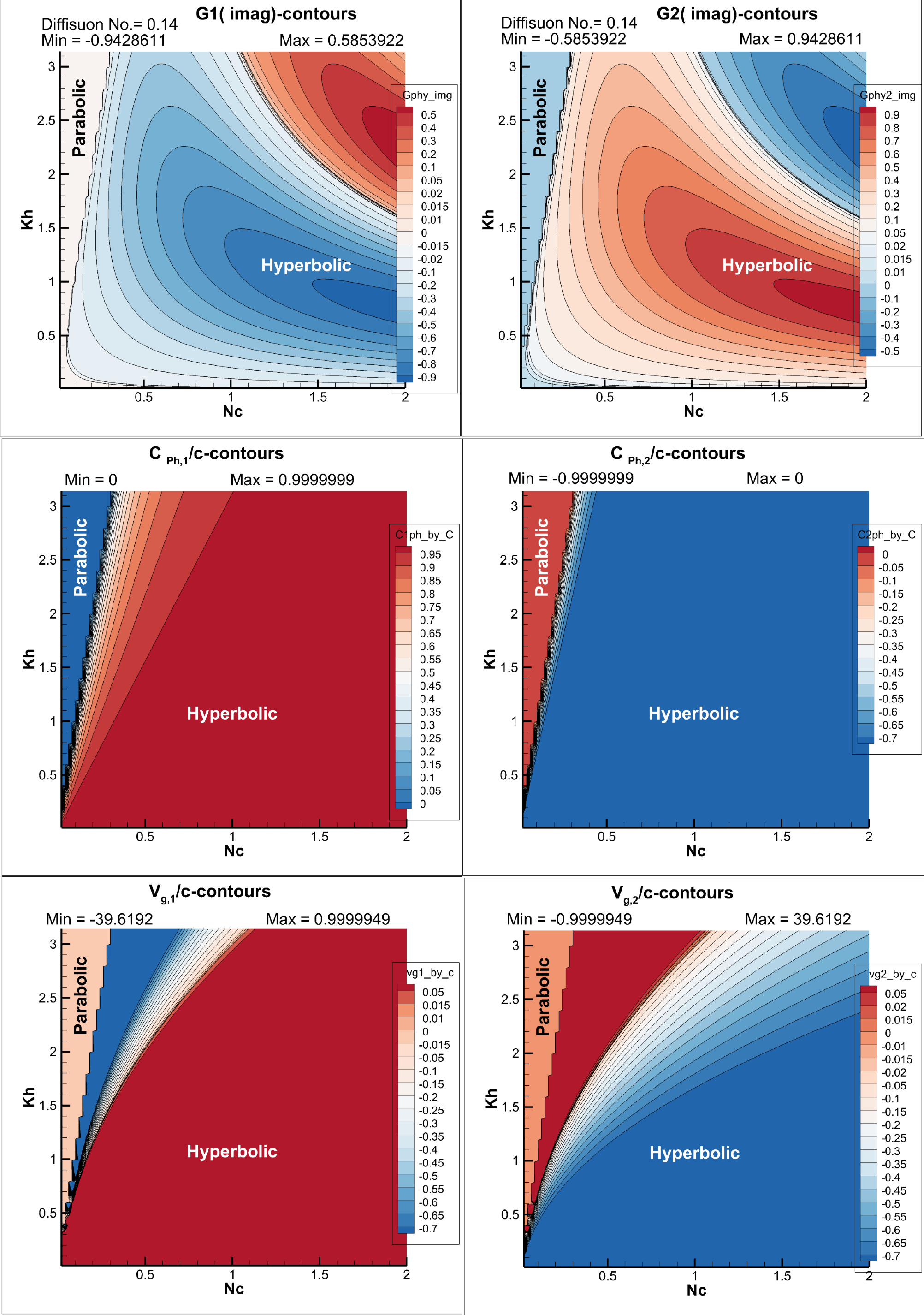}
\caption{The top, middle and bottom frames show both the modes of imaginary part of amplification factor, phase speed and group velocity for the diffusion number of $0.14$.}
\label{Fig1}
\end{figure*}

\subsection{Characteristics of Perturbation Pressure Equation}
The above mentioned acoustic equation derived in the dissipative medium is distinctly different from the one derived by Feynman \cite{Feynman65, Feynman69} for loss-less medium ($\nu_l \equiv 0$). Equation~\eqref{Eq:12} without loss term ($\nu_l =0$) is the classical wave equation as described before \cite{Whitham74,DAlembert1750,Sengupta13,SenguptaBhum20,maxwell1865,maxwell54,Mullothetal2015}. In contrast to the wave equation (being a hyperbolic PDE), the present acoustic equation needs mathematical characterization. As the governing conservation equation is the linearized, compressible Navier-Stokes equation, this is adequately investigated by the global spectral analysis \cite{Sengupta13, SumanSengPrasMoha17}. 

To elucidate the fundamentals, attention is focused here in Eq.~\eqref{Eq:12} for the one-dimensional version of the acoustic equation \cite{Blackstock2000} given by, 
\begin{equation} \label{Eq:13}
\frac{\partial^2 p'}{\partial t^2}  - c^2\frac{\partial^2 p'}{\partial x^2} - \nu_l \frac{\partial^3  p' }{\partial t\partial x^2}  = 0
\end{equation}

The hydrodynamic and acoustic events occur on disparate scales, and it remains as a challenge to simultaneously solve flow and acoustic problems. For the purpose of analysis, represent the fluctuating pressure by, 
\begin{equation} \label{Eq:14}
p'(x,t) = \int \int \hat p (k,\omega)e^{ i (kx-\omega t)} dkdw
\end{equation}
Rewriting Eq.\eqref{Eq:13} in the spectral plane by using the above representation, one gets the quadratic dispersion relation as, 
\begin{equation} \label{Eq:15}
\omega ^2  + i \nu_l k^2 \omega - c^2 k^2 = 0 
\end{equation}
This yields the dispersion relations for the two components of the solution as 
\begin{equation}\label{Eq:16}
\omega_{1,2} = \frac{-i \nu_l k^2}{2} \pm kcf
\end{equation}
\noindent where we denote, $f = \sqrt{1-\left(\frac{\nu_l k}{2c} \right)^2}$. Treating the wavenumber $k$, as the independent variable, the dispersion relation in Eq.~\eqref{Eq:16}, provides the following amplification factors as \cite{Sengupta13, SumanSengPrasMoha17}, 
\begin{equation}
G_{1,2} = e^{-i \omega_{1,2} T_s}
\end{equation}
for the introduced time scale $T_s$, and for $f > 0$, these complex exponents indicate phase shifts given by, 
\begin{equation}\label{Eq:19}
\beta_{1,2} = \pm kcf~T_s
\end{equation}
Thus, the positive value of wavenumber-dependent $f$ indicates the dispersive action of the  dissipative medium, in contrast to the non-dispersive nature of the classical wave equation, Eq.~\eqref{Eq:1}. In general, the phase speed, $c_{ph}$ \cite{Sengupta21} and the phase shifts $\beta$ are related by \cite{Sengupta13}, 

\begin{equation}
    \beta = k c_{ph} T_s, 
\end{equation}
which gives the nondimensional phase speeds of the perturbation equation as, 
\begin{equation} \label{Eq:21}
    \frac{c_{ph 1,2}}{c} = \frac{\beta_{1,2}}{kc T_s} = \pm f 
\end{equation}

The corresponding group velocity components ($v_{g 1,2}$) as given in the literature\cite{Sengupta21} of the perturbation equation are obtained from Eq.~\eqref{Eq:16} as, 

\begin{equation} \label{Eq:22}
    v_{g 1,2} = \frac{d \omega_{1,2}}{dk} = \pm cf \mp \frac{\left(k \nu_l \right)^2}{4fc} - i \nu_l k
\end{equation}
The imaginary part in the above equation is not present for a physical system that does not admit anti-diffusion \cite{SumanSengPrasMoha17}. Here, the diffusion number ($P_e$) is introduced as $\frac{\nu_l~T_S}{L_s^2}$, with $L_s$ taken as a length scale.

In Fig. 1, some typical theoretical results are presented for a single diffusion number case of $P_e = 0.14$, with both modes depicted. In the top frames, the imaginary part of $G_{1,2}$ is shown for this $P_e$ only in ($kL_s,N_c$)-plane, where the nondimensional phase speed (in the absence of losses) is given by, $N_c = c T_s/ L_s$. This is an important feature of $f$, that introduces a cut-off wavenumber ($k_c$) for $f=0$, i.e. $k_c=\frac{2c}{\nu_l}$. For $k>k_c$, the dispersion relation will be strictly imaginary. Above $k_c$, the absence of this imaginary part of $G_{1,2}$ renders the amplification factors to be strictly diffusive, which is representative of a parabolic PDE. In the remaining part of ($kL_s,N_c$)-plane, the perturbation equation represents an attenuated wave, typically representing a hyperbolic PDE.

In the middle two frames, the two components of the physical phase speed given in Eq.~\eqref{Eq:21} are plotted in the same ($kL_s,N_c$)-plane for $P_e=0.14$, and the left triangular portion has the contour value equal to zero, once again implying the parabolic nature of the governing perturbation equation there, as this feature is noted for both the modes. In the hyperbolic part of the domains in ($kL_s,N_c$)-plane, the red region indicates right-running wave and the blue contours indicate left-running dissipative wave. This anisotropy can be noted in Eq.~\eqref{Eq:16} for $k>k_c$, by rewriting the values of $G_1~and~ G_2$ as given by,\\
$(G_1)=e^{-\frac{ P_e (kL_s)^2}{2}}e^{kL_s N_c\sqrt{|1-{(\frac{k}{k_c})}^2}|}$\\
$(G_2)=e^{-\frac{ P_e (kL_s)^2}{2}}e^{-kL_s N_c\sqrt{|1-{(\frac{k}{k_c})}^2}|}$\\

In the bottom two frames of fig..1, the two components of the physical group velocity given in Eq.~\eqref{Eq:22} are plotted in the same ($kL_s,N_c$)-plane for $P_e=0.14$, and the extreme left triangular portion has the contour value equal to zero, once again implying the parabolic nature of the governing perturbation equation, noted for both the modes carrying energy. In the hyperbolic part of the domains in ($kh,N_c$)-plane, the red region indicates right-running wave and the blue contours indicate left-running wave. The boundary between the parabolic and hyperbolic PDEs are defined from Eq.~\eqref{Eq:16} for which $f=0$, rendering $\omega_{1,2}$ as strictly imaginary.

Apart from the dispersive nature of the perturbation equation, another aspect of this equation is in propagating the fluctuating pressure for different wavenumbers, as described next.  

\subsection{Wavenumber dependence of perturbation equation}

One of the central results of the perturbation equation is the dispersive nature of the  dissipative medium for the propagation of fluctuating pressure, which is evident from Eq.~\eqref{Eq:16}. One also notes the ultraviolet range of $k \rightarrow \infty$, when one can approximate, $f \approx i |f|$ and then $\omega_{1,2}$ become purely imaginary, and that can explain the absence of ultraviolet catastrophes (that have been explained for electromagnetic radiation \cite{ehrenfest1911zuge}) for real flows with viscous contributions present. 

However, there is another possibility of a qualitative change in the characteristics of the perturbation equation itself. One notices a cut-off wavenumber ($k_c = c(\lambda + 2\mu)/(2\bar \rho h^2)$) for $f\equiv 0$. Above this $k_c$, $\omega_{1,2}$ become strictly imaginary, as explained above. This is a novel result that shows the small perturbations in fluid flows and in aero-acoustics to be given by the fluctuating pressure which displays damped wavy nature for wavenumbers lower than $k_c$, and for the wavenumbers above $k_c$, the perturbation pressure becomes strictly diffusive. Thus, the value of $k_c$ demarcates the wavenumber, above which the mechanical energy will be fully converted to heat - a conjecture that is often used to define the Kolmogorov's scale \cite{MoninYagl71} in turbulent flows.  

An early preliminary version of this research can be found in Sengupta et al.~\cite{arxiv2022} for the one-dimensional perturbation field. In the following, we discuss the dispersion relation of multi-dimensional perturbation field.

\section{Dispersion Relation for multi-dimensional perturbation field}
Here, following the vector calculus notations we present the perturbation momentum equation as,\\
\begin{equation} \label{eqn:01}
\bar \rho  \frac{\partial \vec V '}{\partial t}  = -\nabla p' + \mathcal{S}
\end{equation}

where,
\begin{equation}\label{eqn:02}
    \mathcal{S} = \nabla \cdot  \left\{ \lambda \left(\nabla \cdot \vec V'  \right)\bf{I} +  \mu \left( \nabla \vec V' + \nabla \vec V^{' T} \right) \right\}
\end{equation}

When $\lambda$ and $\mu$ are considered constant, and the vector identities  $\nabla \cdot \left(\nabla \cdot \vec V'\right) \bf{I} =
\nabla \cdot \nabla \vec V'^T =  \nabla\left(\nabla \cdot \vec V'\right) $ and $\nabla \cdot \nabla \vec V' = \nabla^2 \vec V'$ help us to reduce $\mathcal{S}$ as,

\begin{equation}\label{eqn:02a}
    \mathcal{S} =  \left( \lambda + \mu \right) \nabla\left(\nabla \cdot \vec V'\right) +  \mu \nabla^2 \vec V'
\end{equation}

Taking the time derivative of Eq.~\eqref{eqn:01} one gets,

\begin{equation} \label{eqn:03}
\bar \rho  \frac{\partial^2 \vec V '}{\partial t^2}  = -\nabla \frac{\partial p'}{ \partial t} + \frac{\partial \mathcal{S}}{ \partial t}
\end{equation}

Furthermore, using the polytropic relation one gets,

\begin{equation} \label{eqn:04}
    \bar \rho  \frac{\partial^2 \vec V '}{\partial t^2}  = -c^2 \nabla \frac{\partial \rho'}{ \partial t} + \frac{\partial \mathcal{S}}{ \partial t}
\end{equation}

Also, using the continuity equation, one gets

\begin{equation} \label{eqn:05}
    \frac{\partial^2 \vec V '}{\partial t^2}  = c^2 \nabla \left( \nabla \cdot \vec V' \right)  + \frac{1}{\bar \rho}\frac{\partial \mathcal{S}}{ \partial t}
\end{equation}
or\\
\begin{equation} \label{eqn:06}
    \frac{\partial^2 \vec V '}{\partial t^2}  = c^2 \nabla \left( \nabla \cdot \vec V' \right)  +  \frac{1}{\bar \rho} \frac{\partial}{ \partial t} \left[ \left( \lambda + \mu \right) \nabla\left(\nabla \cdot \vec V'\right) +  \mu \nabla^2 \vec V' \right]
\end{equation}
or\\
\begin{equation} \label{eqn:07}
    \frac{\partial^2 \vec V '}{\partial t^2}  =  \left(c^2  + \frac{\lambda + \mu}{\bar \rho} \frac{\partial}{ \partial t} \right) \nabla \left( \nabla \cdot \vec V' \right) + \frac{\mu}{\bar \rho} \frac{\partial}{ \partial t} \nabla^2 \vec V'
\end{equation}

The ($x$, $y$, $z$) components of the above equation can be written as,
\begin{eqnarray}
    \frac{\partial^2 u'}{\partial t^2}  &=&  \left(c^2  + \frac{\lambda + \mu}{\bar \rho} \frac{\partial}{ \partial t} \right) \frac{\partial}{ \partial x} \left[\frac{\partial u'}{\partial x} + \frac{\partial v'}{\partial y} + \frac{\partial w'}{\partial z} \right] + \frac{\mu}{\bar \rho} \frac{\partial}{ \partial t} \nabla^2 u' \nonumber \\
    \frac{\partial^2 v'}{\partial t^2}  &=&  \left(c^2  + \frac{\lambda + \mu}{\bar \rho} \frac{\partial}{ \partial t} \right) \frac{\partial}{ \partial y} \left[\frac{\partial u'}{\partial x} + \frac{\partial v'}{\partial y} + \frac{\partial w'}{\partial z} \right] + \frac{\mu}{\bar \rho} \frac{\partial}{ \partial t} \nabla^2 v' \nonumber \\
    \frac{\partial^2 w'}{\partial t^2}  &=&  \left(c^2  + \frac{\lambda + \mu}{\bar \rho} \frac{\partial}{ \partial t} \right) \frac{\partial}{ \partial z} \left[\frac{\partial u'}{\partial x} + \frac{\partial v'}{\partial y} + \frac{\partial w'}{\partial z} \right] + \frac{\mu}{\bar \rho} \frac{\partial}{ \partial t} \nabla^2 w' \label{eqn:08}
\end{eqnarray}
It is to be noted that the $x$-component of the equation for a one-dimensional problem can be written as,

\begin{equation}
    \frac{\partial^2 u'}{\partial t^2}  = c^2 \frac{\partial^2 u'}{ \partial^2 x} + \frac{\lambda + 2\mu}{\bar \rho} \frac{\partial^3 u'}{ \partial t\partial^2 x}
\end{equation}

This indicates that the perturbation velocity and pressure fields are given by the identical equation in one-dimension, which is not the case in multiple dimension.\\
The Cartesian components of velocity vector can be represented in the spectral plane to get the dispersion relation in multiple dimensions as,

\begin{eqnarray}
    u'(\vec r,t) = \int \int \hat u ( \vec k,\omega)e^{i (\vec k \cdot \vec r-\omega t)} d\vec k~d\omega \\
    v'(\vec r,t) = \int \int \hat v ( \vec k,\omega)e^{i (\vec k \cdot \vec r-\omega t)} d\vec k~d\omega \\
    w'(\vec r,t) = \int \int \hat w ( \vec k,\omega)e^{i (\vec k \cdot \vec r-\omega t)} d\vec k~d\omega
\end{eqnarray}
Here, the physical space-time domain ($\vec r$ and $t$) is mapped in the circular frequency ($\omega$) and wavenumber vector ($\vec k$) plane.\\
Thus, Eq.~\eqref{eqn:08} gives,

\begin{eqnarray}
    \omega^2 \hat u &-& k_x \left( c^2 - i \omega \frac{\lambda + \mu}{\bar \rho} \right) \left( k_x \hat u + k_y \hat v + k_z \hat w \right) + \frac{i \omega \mu}{\bar \rho}  \left(k_x^2 + k_y^2 + k_z^2 \right) \hat u = 0 \nonumber \\
    \omega^2 \hat v &-& k_y \left( c^2 - i \omega \frac{\lambda + \mu}{\bar \rho} \right) \left( k_x \hat u + k_y \hat v + k_z \hat w \right) + \frac{i \omega \mu}{\bar \rho}  \left(k_x^2 + k_y^2 + k_z^2 \right) \hat v = 0 \nonumber \\
    \omega^2 \hat w &-& k_z \left( c^2 - i \omega \frac{\lambda + \mu}{\bar \rho} \right) \left( k_x \hat u + k_y \hat v + k_z \hat w \right) + \frac{i \omega \mu}{\bar \rho}  \left(k_x^2 + k_y^2 + k_z^2 \right) \hat w = 0
\end{eqnarray}

If we let, $\mathcal{P} =  i \frac{\lambda + \mu}{\bar \rho}$ and  $\mathcal{B} = \frac{i \mu}{\bar \rho}  \left(k_x^2 + k_y^2 + k_z^2 \right)$, then Eq.~\eqref{eqn:35} can be rewitten as
%
%
%
%
%
\begin{equation}
\begin{bmatrix}
\omega^2 + k_x^2 \left(\omega \mathcal{P}-c^2\right) + \mathcal{B}\omega & k_xk_y \mathcal{B}\omega & k_xk_z \mathcal{B}\omega \\
k_xk_y \mathcal{B}\omega & \omega^2 + k_y^2 \left(\omega \mathcal{P}-c^2\right) + \mathcal{B}\omega & k_yk_z \mathcal{B}\omega \\
k_xk_z \mathcal{B}\omega & k_yk_z \mathcal{B}\omega & \omega^2  + k_z^2 \left(\omega \mathcal{P}-c^2\right) + \mathcal{B}\omega
\end{bmatrix}
\begin{bmatrix}
\hat u \\ \hat v \\ \hat w
\end{bmatrix}
=
\begin{bmatrix}
0 \\ 0\\ 0
\end{bmatrix}
\end{equation}

By equating the determinant of the above matrix to zero, we get the dispersion relation as,
\begin{eqnarray}
\omega^{6} && + \omega^{5} \left(\mathcal{P} k_x^{2} + \mathcal{P} k_y^{2} + \mathcal{P} k_z^{2} + 3 \mathcal{B} \right)  \nonumber \\
&& + \omega^{4} \left(\mathcal{P}^{2} k_x^{2} k_y^{2} + \mathcal{P}^{2} k_x^{2} k_z^{2} + \mathcal{P}^{2} k_y^{2} k_z^{2} + 2 \mathcal{P} \mathcal{B} k_x^{2} + 2 \mathcal{P} \mathcal{B} k_y^{2} + 2 \mathcal{P} \mathcal{B} k_z^{2} - \mathcal{B}^{2} k_x^{2} k_y^{2}\right) + \nonumber \\
&& + \omega^{4} \left(  - \mathcal{B}^{2} k_x^{2} k_z^{2}  - \mathcal{B}^{2} k_y^{2} k_z^{2} + 3 \mathcal{B}^{2} - k_x^{2} c^{2} - k_y^{2} c^{2} - k_z^{2} c^{2}\right) \nonumber \\
&& + \omega^{3} \left(\mathcal{P}^{3} k_x^{2} k_y^{2} k_z^{2} + \mathcal{P}^{2} \mathcal{B} k_x^{2} k_y^{2} + \mathcal{P}^{2} \mathcal{B} k_x^{2} k_z^{2} + \mathcal{P}^{2} \mathcal{B} k_y^{2} k_z^{2} - 3 \mathcal{P} \mathcal{B}^{2} k_x^{2} k_y^{2} k_z^{2} \right) \nonumber \\
&& + \omega^{3} \left(\mathcal{P} \mathcal{B}^{2} k_x^{2} + \mathcal{P} \mathcal{B}^{2} k_y^{2} + \mathcal{P} \mathcal{B}^{2} k_z^{2} - 2 \mathcal{P} k_x^{2} k_y^{2} c^{2} - 2 \mathcal{P} k_x^{2} k_z^{2} c^{2} - 2 \mathcal{P} k_y^{2} k_z^{2} c^{2} + 2 \mathcal{B}^{3} k_x^{2} k_y^{2} k_z^{2} \right) \nonumber \\
&& + \omega^{3} \left(- \mathcal{B}^{3} k_x^{2} k_y^{2} - \mathcal{B}^{3} k_x^{2} k_z^{2} - \mathcal{B}^{3} k_y^{2} k_z^{2} + \mathcal{B}^{3} - 2 \mathcal{B} k_x^{2} c^{2} - 2 \mathcal{B} k_y^{2} c^{2} - 2 \mathcal{B} k_z^{2} c^{2}\right)  \nonumber \\
&& + \omega^{2} \left(- 3 \mathcal{P}^{2} k_x^{2} k_y^{2} k_z^{2} c^{2} - 2 \mathcal{P} \mathcal{B} k_x^{2} k_y^{2} c^{2} - 2 \mathcal{P} \mathcal{B} k_x^{2} k_z^{2} c^{2} - 2 \mathcal{P} \mathcal{B} k_y^{2} k_z^{2} c^{2} + 3 \mathcal{B}^{2} k_x^{2} k_y^{2} k_z^{2} c^{2} \right)  \nonumber \\
&& + \omega^{2} \left(- \mathcal{B}^{2} k_x^{2} c^{2} - \mathcal{B}^{2} k_y^{2} c^{2} - \mathcal{B}^{2} k_z^{2} c^{2} + k_x^{2} k_y^{2} c^{4} + k_x^{2} k_z^{2} c^{4} + k_y^{2} k_z^{2} c^{4}\right)   \nonumber \\
&& + \omega \left(3 \mathcal{P} k_x^{2} k_y^{2} k_z^{2} c^{4} + \mathcal{B} k_x^{2} k_y^{2} c^{4} + \mathcal{B} k_x^{2} k_z^{2} c^{4} + \mathcal{B} k_y^{2} k_z^{2} c^{4}\right) - k_x^{2} k_y^{2} k_z^{2} c^{6} = 0
\end{eqnarray}
This is the dispersion relation for multi-dimensional perturbation field in a quiescent medium which can be solved only numerically.

\section*{AUTHOR DECLARATIONS}
The authors are grateful to Prof. A. Sengupta for the useful discussion.
\subsection*{Conflict of Interest}
The authors have no conflicts to disclose.

\section*{DATA AVAILABILITY}
The data that support the findings of this study are available from the corresponding author upon reasonable request.

\bibliography{acoustics}
\end{document}